\documentclass[pra,twocolumn,preprintnumbers,amsmath,amssymb]{revtex4}

\usepackage[T1]{fontenc}
\usepackage[ansinew]{inputenc}
\usepackage[T1]{fontenc}
\usepackage{ae,aecompl}
\usepackage[english]{babel}
\usepackage[dvips]{graphicx}
\usepackage{amsmath}
\usepackage{enumerate,amsthm,amsmath,amssymb,color,
float}
\usepackage{natbib}
\usepackage{epstopdf}
\usepackage{pifont}
\usepackage{eufrak}
\newcommand{\braket}[2]{\mbox{$\langle #1|#2\rangle$}}
\newcommand{\ketbra}[2]{\mbox{$|#1\rangle\langle #2|$}}
\newcommand{\op}[1]{\mbox{\boldmath $\hat{#1}$}}

\newcommand{\ket}[1]{\vert#1\rangle}
\newcommand{\bra}[1]{\langle#1\vert}

\begin{document}

\title{A method for characterizing coherent-state quantum gates}
\author{R\'emi Blandino$^1$, Franck Ferreyrol$^1$, Marco Barbieri$^1$, Philippe Grangier$^1$ and Rosa Tualle-Brouri$^{1,2}$}
\affiliation{$^1$Groupe d'Optique Quantique, Laboratoire Charles Fabry, Institut d'Optique, CNRS, Universit\'e Paris-Sud, Campus Polytechnique, RD 128, 91127 Palaiseau cedex, France }
\affiliation{$^2$Institut Universitaire de France}
\begin{abstract}
We discuss and implement experimentally a method for characterizing quantum gates operating on superpositions of coherent states. The peculiarity of this encoding of qubits is to work with a non-orthogonal 
basis, and therefore some technical limitations prevent us from using standard methods, such as process tomography. We adopt a different technique, that relies on some a-priori knowledge about the physics 
underlying the functioning of the device. A parameter characterizing the global quality of the quantum gate is obtained by ``virtually'' processing an entangled state. 
\end{abstract}
\maketitle

Future quantum computers will be able to incorporate quantum logic in the treatment of information. For this purpose, operations are performed on quantum systems in order to implement the desired processing 
steps. Unavoidably, such operations will only be an approximation of the ideal quantum gates, and the degree within which this approximation is acceptable is determined by the effectiveness of error correction 
codes \cite{Bill}. The characterization of quantum operations is then an important step to establish 
practical limits in the use of such devices.

A crucial requirement is that the system remains under controlled manipulation, and is well-preserved from coupling to the environment, so that the action of the gates is not spoiled by decoherence. Optics offers 
an interesting option in this sense; within quantum optics, several proposals have been put forward, either based on a discrete \cite{KLM}, or a continuous variable approach \cite{book}.

Recently, a different point of view has emerged, which aims to join the strengths from both worlds, and encodes quantum bits (qubits) in superpositions of weak coherent states \cite{Tim, AlexeiG}. The peculiarity 
of this approach is that the logical states $\ket{0}$ and $\ket{1}$ are represented by two non-orthogonal (thus non mutually-exclusive) states of the system, viz. two coherent states with the same amplitude and 
opposite phase $\ket{\alpha}$ and $\ket{{-}\alpha}$; this has demonstrated to be an error-correctable approach for moderately low intensities $|\alpha|\sim1.5$, where the overlap is ${|\braket{\alpha}{{-}\alpha}|
^2\sim10^{-4}}$ \cite{Austin}. The cost of this choice is that the gates are necessarily probabilistic, with a success rate that depends on this overlap; remarkably, though, the comparison with other optical schemes in 
terms of resources needed for scalability seems favorable \cite{Austin,Haselgrove}.

Gates in a coherent states architecture cannot be described by ordinary quantum operators, as they are defined using non-orthogonal states as a basis. Moreover, decoherence processes most likely will not leave the output state inside the original reduced Hilbert space spanned by $\ket{\alpha}$ and $\ket{{-}\alpha}$. The current technologies do not even warrant that input test states belong to this space. Therefore, we cannot simply use standard techniques, such as process 
tomography \cite{Cirac, Chuang, White, Lvovsky}, for a characterization. 

In this article we propose a way for characterizing these gates, which does not rely on a black-box approach \cite{Chuang}, but requires some modeling of the functioning of the gate. This is a realistic approach, 
since such a knowledge is needed to achieve the desired level of control. One can then identify a small number of parameters, accessible to the experimentalist, by which the gate process can be modeled. 

Our demonstration concerns the $\pi$-phase gate for coherent-state qubits proposed by Marek and Fiur\'a\v sek \cite{Marek}. The operating principle of the gate is that coherent states are eigenstates of the 
photon subtraction operator: $\hat a \ket{\pm\alpha}=\pm\alpha\ket{\pm \alpha}$. This operation thus corresponds to a $\pi$-phase shift, up to an overall constant. In the laboratory, photon subtraction can be 
approximated by a beam splitter with low reflectivity, followed by detection on a photon counter. There have been demonstrations of such an effect \cite{Jerome,Alexei,Valentina,Giappi}, but so far the 
characterization has focused on the states that could be produced by this technique, rather than on the device itself. While it has not been investigated if these gates are sufficient for full scalability, they 
nevertheless provide a reliable way to implement logical operations and test this architecture, in the same vein as other optical realizations \cite{altromio,mio}. Therefore, they are an interesting testbed for our 
method.

\begin{figure}[t]
\includegraphics[width=0.98 \columnwidth ]{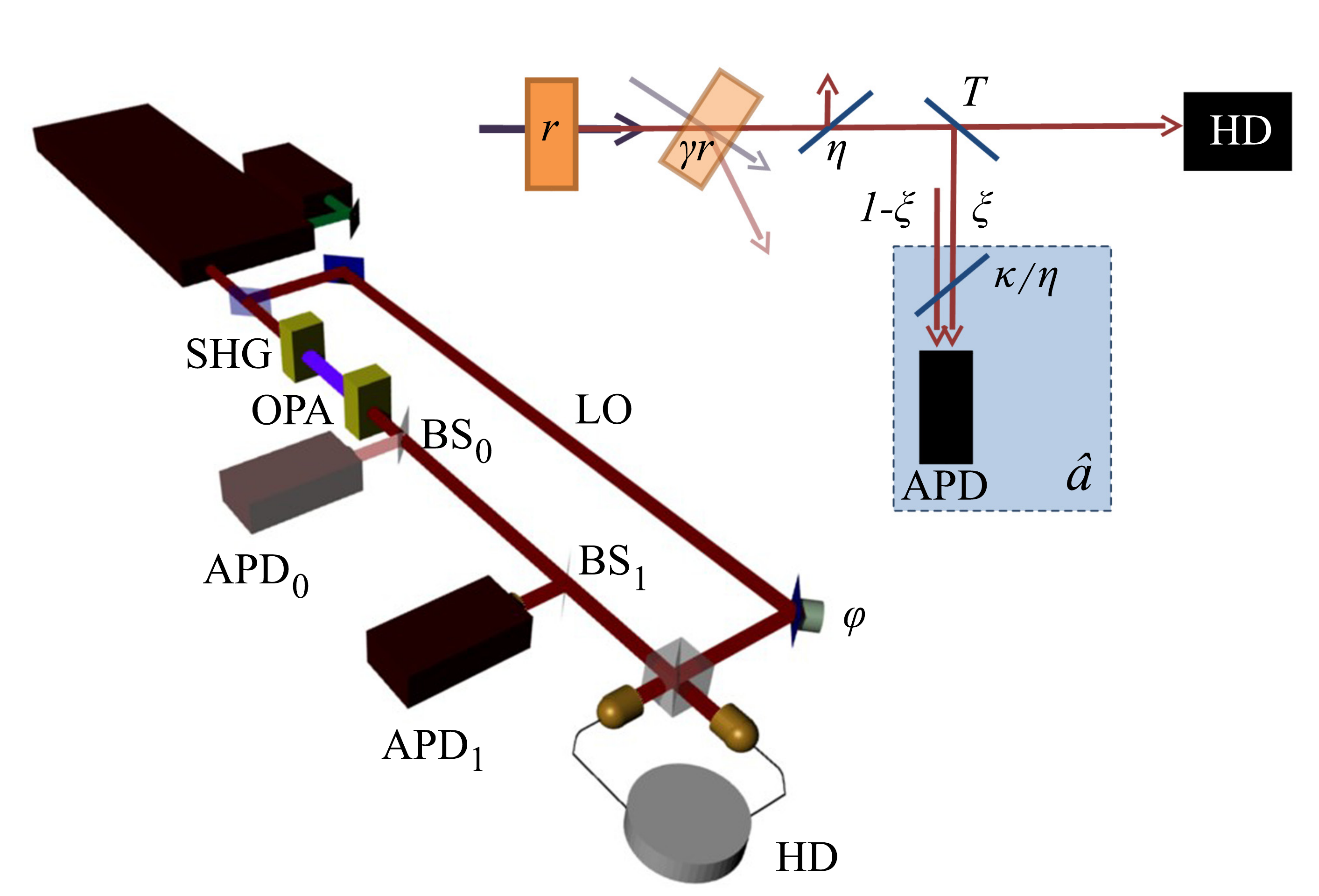}
\caption{Left : scheme of the experiment. The main laser is a pulsed Ti:Sapphire at $\lambda{=}$850 nm, pulsed at a repetition rate 800 kHz, with a pulse width of 200~fs. The light pulses are frequency doubled 
by a
crystal of potassium niobate (SHG) in order to produce a pump beam for the optical parametric amplifier (OPA). This OPA produces a squeezed vacuum, which is then transformed into a squeezed one-photon 
by photon subtraction \cite{sqzphoton}. 
This is performed by directing a small fraction of the beam on the avalanche photodiode $APD_0$ by a beamsplitter $BS_0$, and conditioning the measurement on a detection event. Either the squeezed 
vacuum or the squeezed photon are used as the input of the tested gate, constituted itself by a beamsplitter $BS_1$ and the $APD_1$. Both beamsplitters are realized by a sequence of a half-wave plate and a 
polarizer, so that we could tune both reflectivities to $R=10\%$. This also allow us to take them out of the path of the input for a direct measurement. A small portion of the main laser is used as a local oscillator 
(LO) for a homodyne detector (HD); the phase $\varphi$ is scanned by means of a piezo actuator. \\ Right : Visual representation of the phenomenological model. Symbols are defined in the main text.} 
\label{Fig:setup}
\end{figure}

Testing quantum gates demands to operate them with at least two different orthogonal bases \cite{Holger}. Here the computational basis $\{ \ket{\alpha}, \ket{{-}\alpha}\}$ is trivial, while equal real superpositions 
can be obtained by manipulation of a squeezed vacuum. As shown in Fig.~\ref{Fig:setup}, this state is produced using an optical parametric amplifier (OPA), pumped in the collinear regime. For sufficiently weak 
intensities, the OPA generates a good approximation of an even superposition $\ket{+}{=}\cal{N}_+\left(\ket{\alpha}{+}\ket{{-}\alpha}\right)$, where $\cal{N}_+$ is a normalizing factor \cite{norm}. The odd superposition $\ket{-}{=}
\cal{N}_-\left(\ket{\alpha}{-}\ket{{-}\alpha}\right)$ results from the application of the gate \cite{Alexei}. In our experiment, this is implemented by the beam splitter $BS_0$, and the avalanche photodiode $APD_0$. 
These two states can be used to test the behavior of a second gate constituted by $BS_1$ and $APD_1$.

Although the states we can produce do have marked signatures of the ideal behavior, such as a change in the parity, current technology limits us to some approximations of coherent-state qubits \cite
{Giappi,AlexeiN}, which are not fully suitable for a direct characterization of the gate. In a direct comparison of the observed output to the ideal, it would be hard to deconvolute the errors due to the imperfections 
of the gate, and the imperfections of the input state itself.

These limitations can be relaxed if one concedes to experimentalists some a priori knowledge about the physics of their gate~: a model can then be derived, and used for such a deconvolution, and thus to 
obtain the behavior of the device for ideal inputs. This is somehow similar to the analysis in \cite{Till} to estimate the origin of decoherence in polarization quantum gates. In our case, we can rely on a simple 
phenomenological model \cite{AlexeiN}, which is grounded on a more rigorous multi-mode treatment \cite{Rosa}. 

\begin{figure}[h]
\includegraphics[width=1 \columnwidth]{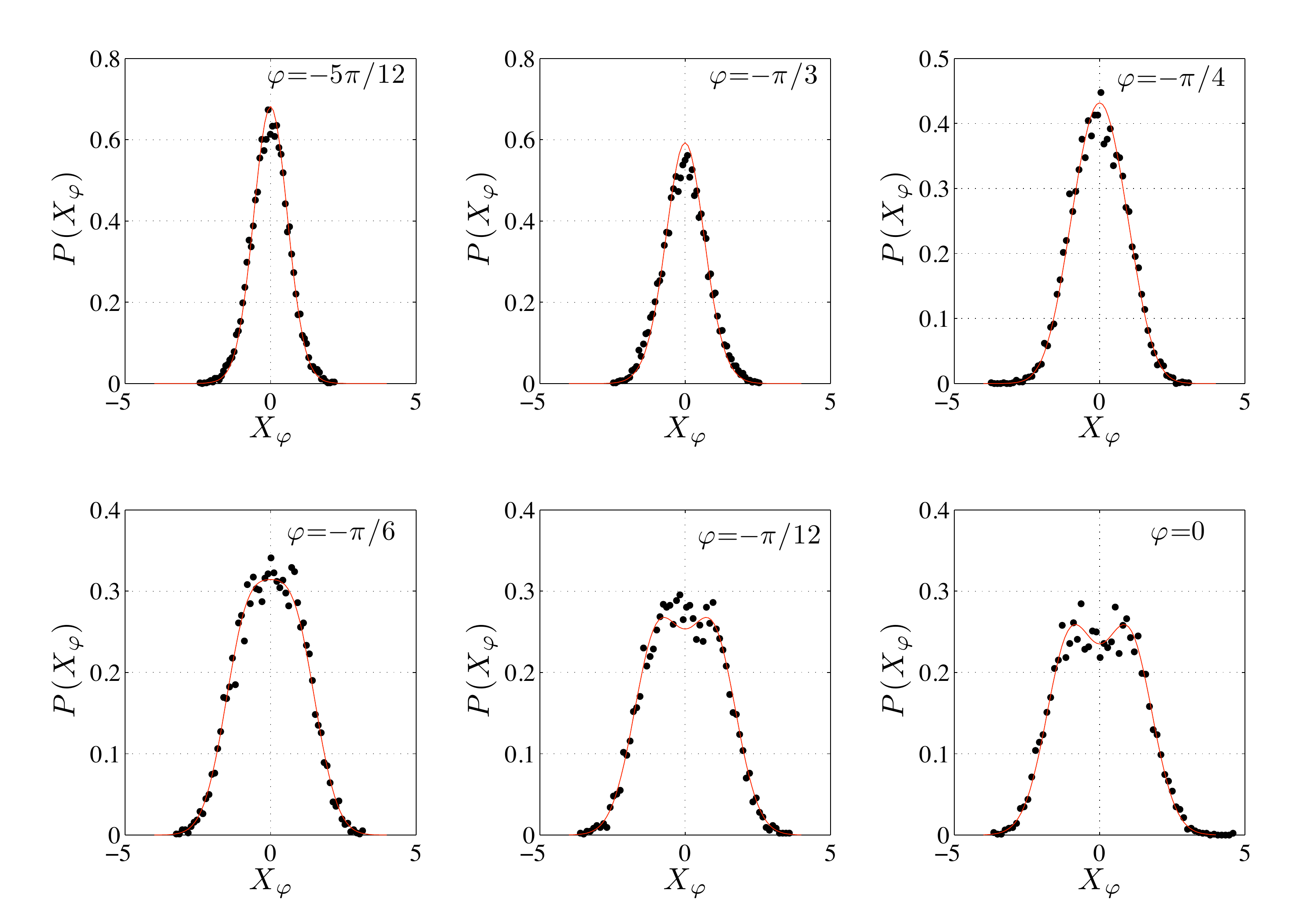}
\caption{Quadrature distributions of a photon-subtracted squeezed photon. The points are the measured values, while the solid lines are the prediction of the model, based on the outcomes from a squeezed 
state. Each histogram is calculated from 10000 quadrature points. We can attribute some noise to fluctuations during the long acquisition time, due to the low two-photon subtraction rate ($\sim$6 events$\cdot$s
$^{-1}$).} 
\label{Fig:histos}
\end{figure}



This model is illustrated in Fig.~\ref{Fig:setup} for the case of single photon subtraction. We consider a perfect squeezer, i.e. a collinear OPA, which can reduce noise by $s{=}e^
{-2r}$ times the shot noise. The process is spoiled by parasite amplification that can be described by a fictive non-collinear OPA injected on one side with the squeezed vacuum; its gain is expressed as $h{=}
\cosh(\gamma r)$. Once we trace out the fictive idler mode, we obtain the following expression for the Wigner function:
$W(x,p) \; { \propto } \; e^{-\frac{x^2}{hs+h-1}} \; e^{-\frac{p^2}{h/s+h-1}}$.

\begin{figure*}[th]
\includegraphics[width= 2.06 \columnwidth]{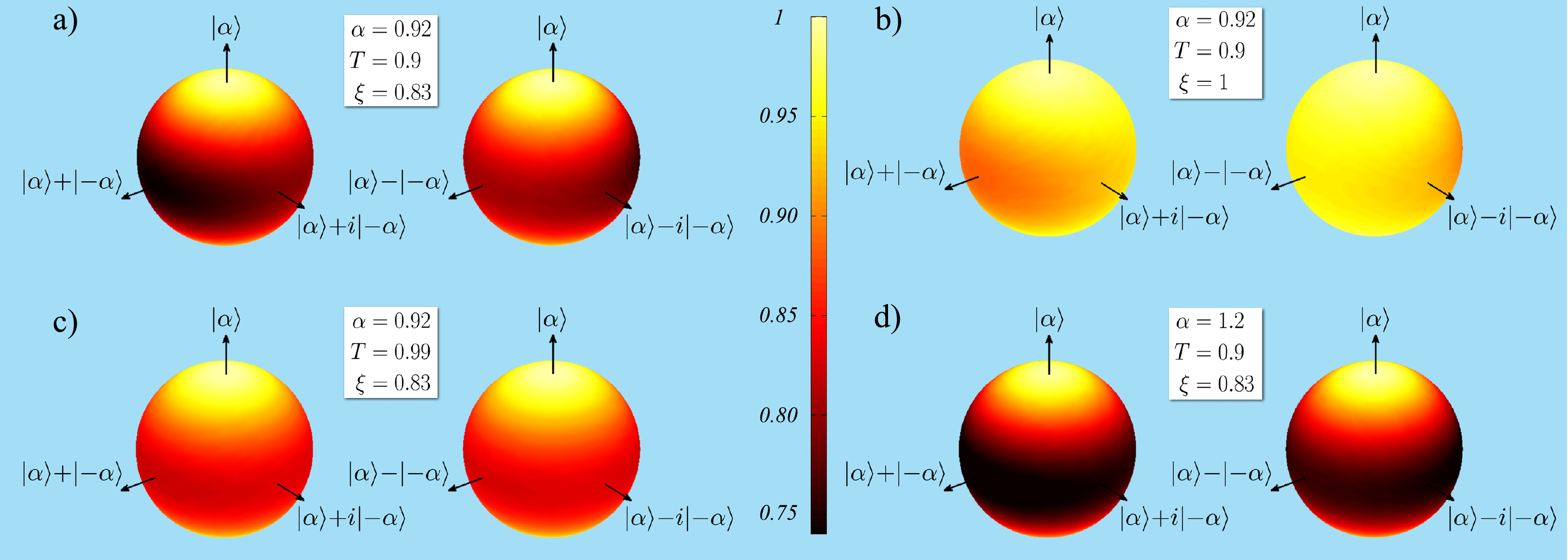}

\caption{Fidelities of the output states for arbitrary ideal inputs. a) for our phase gate; b) for a perfect modal purity $\xi=1$; c) for a device with $T{=}0.99$; d) for our device with $\alpha{=}1.2$. The angles $\theta$ and $\phi$ are the spherical coordinates on the Bloch sphere.} 
\label{Fig:Bloch}
\end{figure*}

This is the squeezed state that is fed in the gate. This consists of a beam-splitter with transmissivity $T{\simeq} 1$ ($T{=}0.9$ in our implementation), and an avalanche photodiode (APD) of efficiency $\kappa$, whose signal is used as 
a trigger event. The mode of the squeezed vacuum is filtered by a single mode fiber and a spectral filter; this operation, though, is performed with finite accuracy. There is then a fraction $\xi$ of the events that 
originates from photon subtraction on the correct mode; the remaining $1{-}\xi$ are actually non-correlated events that corrupt the functioning of the gate. 

The finite efficiency of the homodyne detection (HD) has to be taken into account; for the sake of simplicity, we introduce the homodyne loss $1{-}\eta$ before the beamsplitter rather than before the HD, then we correct the efficiency of the 
APD as $\kappa/\eta$. This has the advantage to simplify the computations and is strictly equivalent to the use of homodyne loss just before the HD and an APD efficiency of $\kappa$. In the limit of small efficiency, we can describe the action of the APD simply as the destruction operator $\hat a$ acting on the reflected mode, that is finally traced out. 

We obtain thus a parametrization of the Wigner function, which depends on a few parameters; some of them are inherent to the state -- the squeezing $s$, the parasite gain $h$ -- other describe the action of the 
gate -- the modal purity $\xi$, the transmissivity $T$ -- and finally the homodyne efficiency $\eta$ describes the detection. These can be either directly measured by a classical signal, as it is the case for $T$, or 
inferred by a best fit on the quadrature histograms. For our analysis, we extrapolate the results for the generated state, by imposing $\eta{\rightarrow}1$. The same modeling is applied when considering the 
double photon subtraction.

For our purpose, it is crucial to note that the functioning of the gate is well described by only two parameters: the reflectivity of the beamsplitter $T$, and the modal purity $\xi$. These parameters can be 
accessed experimentally by using a set of states as a probes. In our case, coherent inputs are not useful, as they are less sensitive to faulty events. We thus use the squeezed vacuum, approximating the even 
cat state, as a probe to estimate the parameters. Then the squeezed photon, approximating the odd cat state, is used to check the consistency of the model. 

More in detail, we performed homodyne measurements on the squeezed vacuum, from which we estimated its density matrix $\rho_0$. This is the starting point of our model, from which we can calculate the 
expected action of the gate, given this state as the input. From a fit, we can derive the value of $\xi$. At this point, we use a squeezed photon as the input; as before its density matrix $\rho_1$ is 
reconstructed by homodyning. The model is now used to estimate the output, fixing $T$ and $\xi$ at the same values as we found in the first case. Finally, we check the consistency of the expected and 
measured outputs.

Typical results are shown in Fig.~\ref{Fig:histos}; we plot six different histograms of quadrature distributions, where the points show the experimental results, and the solid lines the prediction of our model. We 
remark that the parameters are not fitted on the data, but derived from a previous measurements on the squeezed vacuum. This gives us evidence of the reliability of our predictions. 

By using this model, we can estimate the action of the gate on an ideal arbitrary input in the form $\ket{\psi_{\theta,\phi}}{\; =\;}{\cal N}_{\theta,\phi}\left(\cos\frac{\theta}{2}\ket{\alpha}+e^{i\phi}\sin\frac{\theta}{2}\ket{{-}\alpha}\right)$, with $
\theta$ and $\phi$ defining the Bloch sphere of the qubit \cite{Giappi}. For the amplitude $|\alpha|$ of the coherent states we need an estimation of what experimental limits are. For this, we have taken the value 
$|\alpha|{=}0.92$ giving maximal fidelity between our squeezed photon and an ideal odd superposition.

Our analysis is summarized in Fig.~\ref{Fig:Bloch}, where we plot a table of the fidelity $F_{\theta,\phi}$ of the output state with the ideal $\ket{\psi_{-\theta,\phi}}$. The fidelities for our experimental conditions are depicted in Fig.~\ref{Fig:Bloch}a. The first remark concerns the directionality of the device. Better results at the poles with respect to the equator are expected, due to the fact that the high transmission make coherent states largely 
insensitive to the modal purity $\xi$. Furthermore, our analysis shows that odd superpositions work better than the even ones. The limit $\xi{=}1$ (Fig.~\ref{Fig:Bloch}b) reveals that it is an intrinsic feature of the gate: it results from the facts 
that our APD has a low efficiency and to a lesser extent that it can't resolve the photon number. However, as the transmissivity $T$ increases (Fig.~\ref{Fig:Bloch}c), $F_{\theta,\phi}$ also increases and becomes more regular with respect to $\phi$. Indeed, the probability to have several photons reflected to the APD decreases when $T$ increases, and therefore the fact that the APD does not resolve the number of photons becomes less important. For superpositions with $\theta{=}\pi/4$ the limit value of $0.83$ (when $T{\rightarrow}1$) is almost reached for $T{=}0.99$. The downside is that the probability of success of the gate decreases proportionally to $(1{-}T)$. Finally, we see that when $\alpha$ increases (Fig.~\ref{Fig:Bloch}d), $F_{\theta,\phi}$ also becomes more regular with respect to $\phi$, but it reaches lower values, as the probability to have several photons sent to the APD is higher. 

As a more general comment, we conclude that, although the value $\xi{=}0.83$ might be considered satisfactory in our experiment, the operation of the gate is heavily 
influenced by the modal purity. This sets a strong requirement for realizing such gates. 

The results in Fig.~\ref{Fig:Bloch} provide us with extensive information about our device, but fail in delivering us a conclusive answer on how good is the gate overall. As mentioned before, associating a 
quantum process to this operation is non trivial, so we can not formally use Jamio\l kowski's isomorphism and derive a process matrix \cite{Jam}. However, we can still retain the underlying physical idea behind 
the isomorphism. 

Let us briefly recall Jamio\l kowsi's construction; given a process ${\cal E}$ on a space of dimension $d$, we can associate univocally a matrix by considering a maximal entangled state $\ket{\Psi}{=}\frac{1}{\sqrt{d}}\sum_{j{=}1}
^d\ket{j}\ket{j}$, for a given choice of basis $\{\ket{j}\}_{1{\leq}j{\leq}d}$ of the system. The process matrix is then obtained through the application of the process $\mathcal{E}$ to one of the modes of the entangled state, what can be formally written as $\chi{=}I\otimes{\cal E}\left(\ketbra{\Psi}{\Psi}\right)$, that is a state in a larger Hilbert 
space. When a distance between two processes is needed, we can then rely on the measures holding for states in the extended space; in particular, the fidelity between two processes is given by the fidelity 
between the corresponding states: $F({\cal E}_1,{\cal E}_2){=}F({\chi}_1,{\chi}_2)$ \cite{White, mio, Nathan}.

This is in some sense an application of quantum parallelism. Since we are interested in the overall behavior of the gate, we need to estimate its action on all the inputs at the same time: this amounts to feed in the gate 
half of an entangled pair. In light of these considerations, we can adopt as a reasonable figure of merit a fidelity between entangled states. Here we will use as a probe the state $\ket{\Phi^+}{=}\frac{1}{\sqrt{2}}\left(\ket{+}\ket{+}
{+}\ket{-}\ket{-}\right)$, but actually the result does not depend on this choice (see Appendix). 

\begin{figure}
\includegraphics[width=1 \columnwidth, height=6cm]{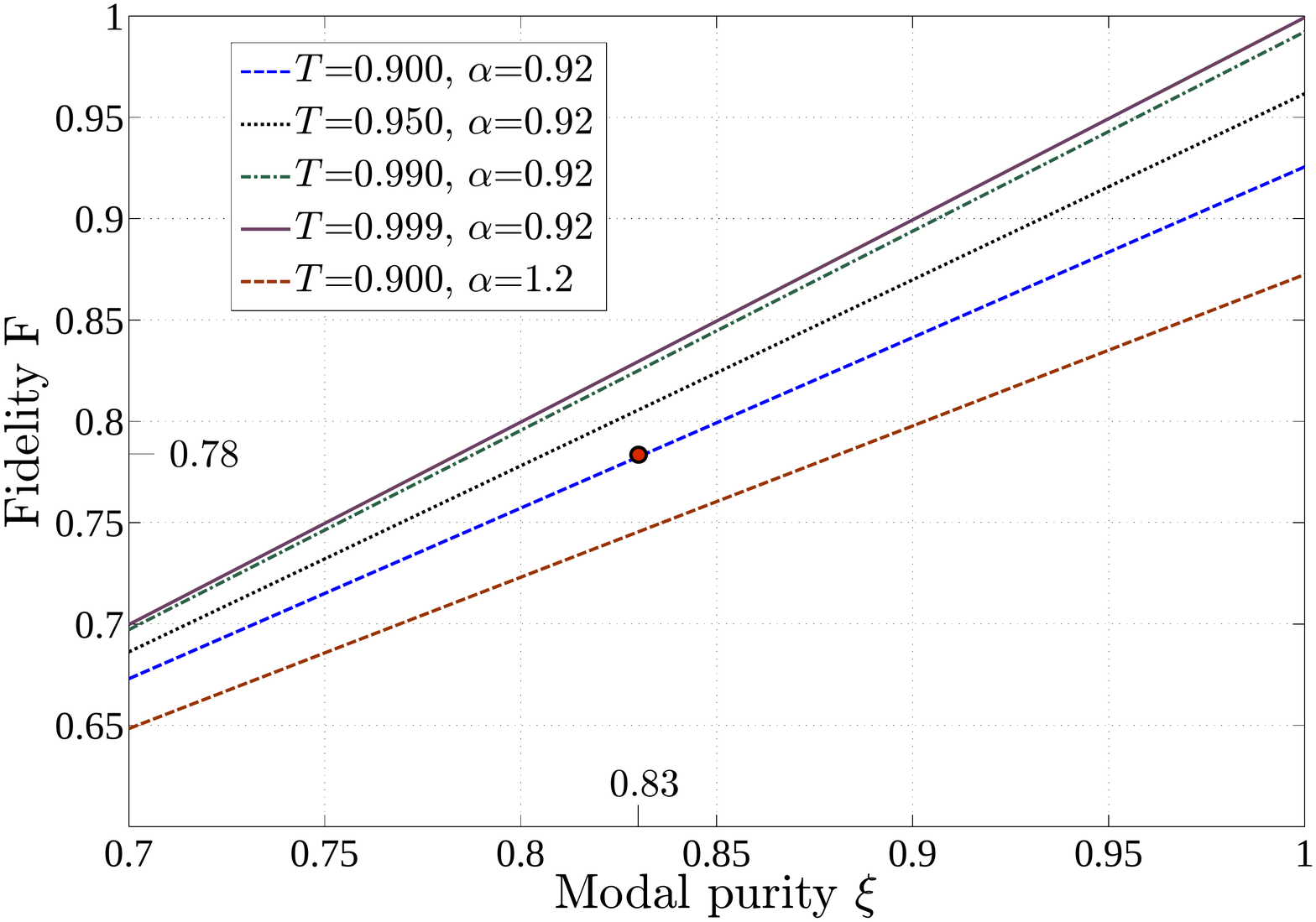}
\caption{Fidelity of an entangled output from the phase gate with the ideal entangled state: we can notice a linear increase of the overlap with the modal purity $\xi$. The red dot corresponds to our experimental parameters.} 
\label{Fig:Entangled}
\end{figure}

We then consider the fidelity $F$ between the entangled output as would be produced by our device, and the ideal target state $\ket{\Omega}{=}\op{I}\otimes\hat{a}\ket{\Phi^+}/\| \op{I}\otimes\hat{a}\ket{\Phi^+}\|$. Our results are shown in Fig.~\ref{Fig:Entangled}: the limiting factor of the performances is mostly the worst-case scenario of an even superposition. We find $F{=}0.78$ for the experimentally observed gate ($\xi{=}0.83$).

\begin{figure}
\includegraphics[width=1 \columnwidth, height=5.7cm]{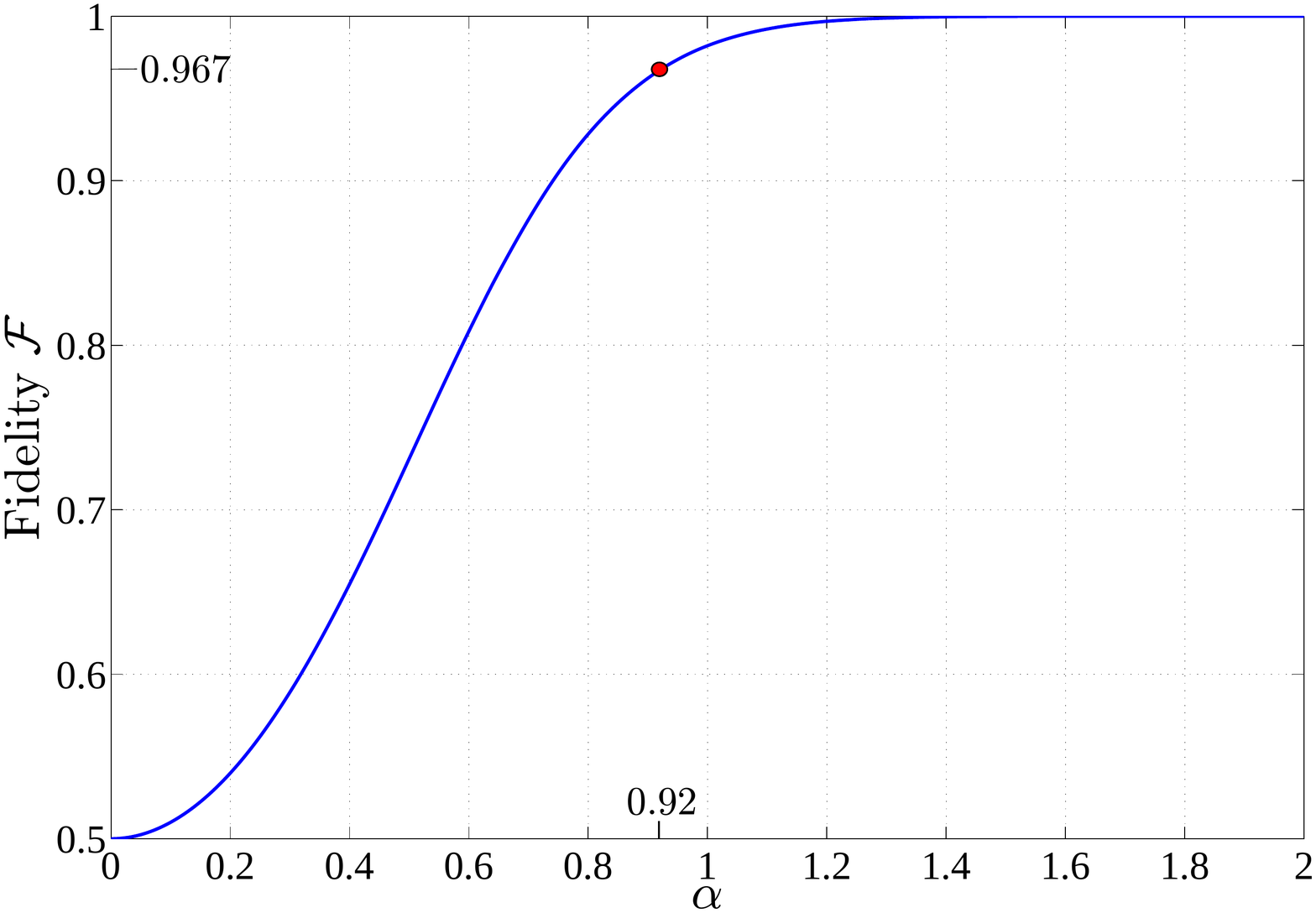}
\caption{Fidelity between $\ket{\Omega}$ and the ideal bit-flipped state $\ket{\Psi^+} $. The red dot corresponds to our experimental parameters.} 
\label{a_gate}
\end{figure}

Let us emphasize that the ideal target state differs from the bit flip state $\ket{\Psi^+}{=}\frac{1}{\sqrt{2}}\left(\ket{+}\ket{-}
{+}\ket{-}\ket{+}\right)$ which would be obtained by an ideal phase gate. This is due to the fact that 
 the states $\ket{\alpha}{\pm}\ket{{-}\alpha}$ have different normalization coefficients $\mathcal{N}_{\pm}$. Such a problem becomes negligible for large enough values of $\alpha$, i.e. when the two states $|{\pm} \alpha \rangle $ become nearly orthogonal. This requirement can be made quantitative by evaluating the fidelity between the target state $|\Omega \rangle$ and $\ket{\Psi^+}$, which is simply :\begin{align}
 \mathcal{F}=|\langle \Omega|\Psi^+\rangle|^2 = \frac{1}{2}(1+\tanh\alpha^2)
 \end{align}This fidelity is plotted on Fig. \ref{a_gate}, and gives an idea of the required values of $|\alpha|$  for the protocol to work correctly, in the sense that cat-state qubit is ``good enough'' for correctly  implementing the desired quantum phase gate. For our value $\alpha{=}0.92$, $\mathcal{F}{=}0.967$.

Summarizing, we have quantified the quality of a simple ($\pi$ phase shift) quantum gate for ``cat-state'' qubits, by evaluating either the fidelity between an arbitrary input state $\ket{\psi_{\theta,\phi}}{\; =\;}{\cal 
N}_{\theta,\phi}\left(\cos\frac{\theta}{2}\ket{\alpha}+e^{i\phi}\sin\frac{\theta}{2}\ket{{-}\alpha}\right)$ and the corresponding output state, or the action of the gate on half an entangled state. This second approach gives a single figure, $F{=}
0.78$ in our case. The method is relatively independent of the quality of the quantum states used to probe the gate, but it requires some a-priori knowledge and modeling of the gate. It appears thus as a stable 
approach for obtaining both a ``quality evaluation" from a single number, and a full description by concatenating simple but efficient theoretical modeling of the gate. 

We thank W.J. Munro for discussion. This work is supported by the EU project COMPAS. F.F. is supported by C’Nano-Ile de France. M.B. is supported by the Marie Curie contract PIEF-GA-2009-236345-PROMETEO.  

\subsection*{Appendix : Invariance of the fidelity with the input entangled state} 
In the main text, we have considered the case when a half of the entangled state $\ket{\Psi^+}{=}\frac{1}{\sqrt{2}}\left(\ket{+}\ket{+}+\ket{-}\ket{-}\right)$ is used as the input of the gate. Here we show that the result does not 
actually depend on the choice of the Bell state. For this purpose, let us consider the general case $\ket{\Psi}{=}\frac{1}{\sqrt{2}}\left(\ket{+}\ket{\mu}+e^{i\phi}\ket{-}\ket{\nu}\right)$, where $\ket{\mu}$ corresponds either to $\ket{+}$ or $\ket{-}$, whereas $\ket{\nu}$ corresponds to the opposite ket: $\ket{\nu}{=}\ket{{-}\mu}$.

The superoperator $\mathcal{E}$ describing the action of the gate can be decomposed in two parts $\mathcal{E}^{(good)}$ and $\mathcal{E}^{(bad)}$ corresponding respectively of the correct and faulty events: $\mathcal{E}{=} \xi \; \mathcal{E}^
{(good)}{+}(1-\xi) \; \mathcal{E}^{(bad)}$.
First, we notice that by linearity we can calculate separately the action of the gate for the correct heralding events and for the faulty ones, and then sum the results with the correct weighting. 

Let us now focus on $\mathcal{E}^{(good)}$, the reasoning being similar for $\mathcal{E}^{(bad)}$. For a given initial state $\rho{=} \sum_{x,y{=}+,-} c_{xy}\ketbra{x}{y}$, the operator $\mathcal{E}^{(good)}$ is non linear. However, as it is modeled with linear operators and partial traces, its non linearity comes only from the final normalization. One can thus write it as a linear operator $\tilde{\mathcal{E}}^{(good)}$ giving a non normalized output state $\tilde{\mathcal{E}}^{(good)}(\rho)$, which is then normalized by its trace: $\mathcal{E}^{(good)}(\rho){=}\tilde{\mathcal{E}}^{(good)}(\rho)/\operatorname{Tr}\{\tilde{\mathcal{E}}^{(good)}(\rho)\}$. We denote with the symbol $\zeta_{xy}$ the action of $\tilde{\mathcal{E}}^{(good)}$ on the operator $\ketbra{x}{y}$. Under the operator $I\otimes \tilde{\mathcal{E}}^{(good)}$ the state $\ketbra{\Psi}{\Psi}$ is transformed into: 

\begin{equation}
\begin{aligned}
&\tilde{\chi}=\frac{1}{2} \big( \ketbra{+}{+}\otimes\zeta_{\mu\mu}+\ketbra{-}{-}\otimes\zeta_{\nu\nu}+\\
&e^{i\phi}\ketbra{-}{+}\otimes\zeta_{\nu\mu}+e^{-i\phi}\ketbra{+}{-}\otimes\zeta_{\mu\nu}\big)
\end{aligned}
\end{equation}

Then, we notice that $\hat{a}\ket{\mu}=\alpha\frac{\mathcal{N}_{\mu}}{\mathcal{N}_{\nu}}\ket{\nu}$. Introducing $c_{\mu\nu}=\frac{\mathcal{N}_{\mu}^2}{\mathcal{N}_{\nu}^2}$, the ideal target state is:
\begin{align}
\ket{\Omega}&=\frac{\op{I}\otimes\hat{a}\ket{\Psi}}{\parallel \op{I}\otimes\hat{a}\ket{\Psi} \parallel} \\
&=\frac{1}{\sqrt{c_{\mu\nu}{+}c_{\nu\mu}}} \big( \sqrt{c_{\mu \nu}}\ket{+}\ket{\nu}+e^{i\phi}\sqrt{c_{\nu \mu}}\ket{-}\ket{\mu} \big)
\end{align} 

We then consider the fidelity between the normalized state $\chi{=}\tilde{\chi}/\text{Tr}\{\tilde{\chi}\}$ and $\ket{\Omega}$. This has the explicit expression: 

\begin{widetext}
\begin{align}
F=\bra{\Omega}\chi\ket{\Omega} =\frac{1}{2\text{Tr}\{\tilde{\chi}\}(c_{\mu\nu}{+}c_{\nu\mu})} \big(c_{\mu\nu} \bra{\nu}\zeta_{\mu\mu}\ket{\nu}+c_{\nu\mu}\bra{\mu}\zeta_{\nu\nu}\ket{\mu}+ \bra{\mu}\zeta_{\nu\mu}\ket{\nu}+\bra{\nu}\zeta_{\mu\nu}\ket{\mu} \big)
\end{align}
\end{widetext}
which does not depend on the choice of $\phi$ and is invariant by exchanging $\mu$ and $\nu$. The same reasoning applies when considering faulty trigger events. In case one would prefer to use the bit-flipped state as the target state, the fidelity would be different, but still independent on the choice of the initial Bell state.


\begin{thebibliography}{33}
\bibitem{norm} The normalization factor ${\cal N}_{\theta,\phi}$ of an arbitrary superposition \mbox{${\cal N}_{\theta,\phi}\left(\cos\frac{\theta}{2}\ket{\alpha}+e^{i\phi}\sin\frac{\theta}{2}\ket{{-}\alpha}\right)$} is equal to \mbox{ $[1+\sin\theta \cos \phi \exp(-2\alpha^2)]^{-1/2}$}. In the particular case of a superposition $\ket{\pm}{=}\mathcal{N}_{\pm}(\ket{\alpha}\pm\ket{{-}\alpha})$, the normalisation factor is: $\mathcal{N}_{\pm}{=}[2(1\pm\exp(-2\alpha^2))]^{-1/2}$.
\bibitem{sqzphoton} Denoting~by~$\hat{S}$~the~squeezing~operator, a photon subtracted squee\-zed state is, up to the normalization: $\hat{a}\hat{S}\ket{0}\propto~\hat{S}\hat{a}^{\dagger}\ket{0}$, which is a squeezed one photon state.
\end{thebibliography}
\end{document}